\documentstyle[aps,epsf,preprint]{revtex}

\begin{document}

\title{
\rightline{LBNL--39652}
\bigskip
Detectability of Strange Matter
in Heavy Ion Experiments}

\author{J\"urgen Schaffner--Bielich$^{a,b,c}$, Carsten Greiner$^d$,
Alexander Diener$^b$ and Horst St\"ocker$^b$}

\address{$a$
The Niels Bohr Institute, Blegdamsvej 17, DK--2100 Copenhagen, Denmark} 
\address{$^b$
Institut f\"ur Theoretische Physik,
J.W. Goethe--Universit\"at,
D--60054 Frankfurt, Germany}
\address{$^c$
Nuclear Science Division, LBNL, Berkeley, 94720 California}
\address{$^d$
Institut f\"ur Theoretische Physik,
Justus--Liebig Universit\"at,
D--35392 Giessen, Germany}
\date{November 22, 1996}
\maketitle

\begin{abstract}
We discuss the properties of two distinct forms of hypothetical strange matter,
small lumps of strange quark matter (strangelets) and of hyperon 
matter (metastable exotic multihypernuclear objects: MEMOs),
with special emphasis on their 
relevance for present and future heavy ion experiments.
The masses of small strangelets up to $A_B=40$ 
are calculated using the MIT bag model with
shell mode filling for various bag parameters.
The strangelets are checked for possible strong and weak hadronic
decays, also taking into account multiple hadron 
decays. It is found that strangelets which are stable against strong decay
are most likely highly negative charged, contrary to previous findings.
Strangelets can be stable against weak
hadronic decay but their masses and charges are still rather high.
This has serious impact on the present high sensitivity searches
in heavy ion experiments at the AGS and CERN facilities.
On the other hand, highly charged MEMOs
are predicted on the basis of
an extended relativistic mean--field model.
Those objects could be detected in future experiments searching for 
short--lived, rare composites. It is demonstrated that
future experiments can be sensitive to 
a much wider variety of strangelets.
\end{abstract}

\section{Introduction}

One of the most fascinating aspects of modern particle physics is the
phase transition from hadronic matter to a deconfined
strong interacting plasma phase. New forms of matter might be 
possible \cite{bodmer} and formed during this transition.
It has been proposed that the introduction of strange quarks into a
plasma with two flavours could lower the Fermi energy of the system 
and thus
the mass of the quark matter. If its mass is lower than the mass of
hadronic (hyperonic) matter with the same strangeness fraction,
it could not decay
(completely)
via strong interactions, which means that it would be metastable
\cite{chin}. If its mass
is even lower than nonstrange nucleonic matter, quark matter would be the
true ground state of nuclear matter \cite{witt84}.

The only way to produce a
quark--gluon plasma in the laboratory are collisions of heavy ions at
ultrarelativistic energies. It has been shown that during the stage of
coexistence of the two phases
the abundantly produced strange and
antistrange quarks are distributed
asymmetrically between the quark and the hadron
phase \cite{grei88,grei91}. The strange quarks are enriched in
the plasma, thus lowering its mass, whereas the antistrange quarks are
found predominantly in the hadronic sector. Radiated
pions and kaons carry away entropy and antistrangeness from the system,
thus cooling it and charging the quark droplet with net strangeness.
This mechanism can
lead to the formation of a droplet of rather cold, strange quark matter,
a strangelet. 

On the other hand, even without a phase
transition strangeness and antistrangeness is abundantly produced in
heavy ion collisions.
Therefore baryonic objects with a high strangeness fraction may be formed,
so--called metastable exotic multihypernuclear objects or
MEMOs \cite{scha92}. They are expected to be bound by energies
up to $E_B/A\approx -22$ MeV and to possess properties quite similar
to those of strangelets \cite{scha93}.

There are several experiments under way at the AGS in
Brookhaven using forward spectroscopy \cite{barr90,aoki92,E878,E886} or an
open geometry \cite{E864} looking for long--lived
charged strangelets ($\tau > 10-100$~ns)
as a `smoking gun' for the formation of a QGP.
The NEWMASS collaboration (experiment NA52) has
recently reported new limits for producing long--lived strangelets 
($\tau > 10^{-6}$ s)
at the higher bombarding energy of the SPS facility at CERN
\cite{bore94}.
The H--dibaryon \cite{jaffe77} 
with quark content ($uuddss$), is possibly the lightest strangelet.
The search for the H--dibaryon in the collisions of heavy ions
opened a very active field of research \cite{DoverH}.
Most recently experiment E888 at AGS 
set new limits on the production of the H--dibaryon
for lifetimes of $\tau > 1$ ns \cite{Hsearch}.
Also quite recently experiment E896 has been approved looking for short--lived
H--dibaryons in forward direction \cite{E896} sensitive to lifetimes of
$\tau\approx 10^{-11}$ s.

Simple coalescence estimates give production probabilities
of strange clusters of the order of $10^{3-A_B-|S|}$, where $S$ denotes the
strangeness and $A_B$ the baryon number of the cluster
\cite{chin,dove90}. 
Hence,
small clusters with $A_B+|S|\le r+3$, where $r$ is the sensitivity of the
apparatus (presently $r \le 12$), are most favoured for detection.
Therefore, if strangelets are formed due to this scenario, baryon numbers
of $A_B \le 12$ are expected.
Dynamical calculations with non--equilibrium
particle emission suggest also $A_B \approx 10-30$ \cite{grei91,bengt} for
initial entropies corresponding to 
AGS and SPS bombarding energies. At higher energy, at RHIC and LHC
colliders, strangelet distillation still works but 
lower mass numbers of $A<10$ are expected \cite{spieles}, which might be
detectable with the ALICE detector at the LHC \cite{alice}.

In the following we examine the properties of both forms of strange
matter for this mass range, its decay properties and its detectability 
for recent and future heavy ion experiments.

\section{Strangelets}
\label{sec2}

In the present investigation, strangelets are treated as noninteracting
fermions within the MIT--Bag model, filling up the bag with exact
single--particle Dirac states following \cite{farhi,grei88}. We point
out that the simple mass formula of Berger and Jaffe \cite{berger} or
Fermi gas models including a curvature term \cite{madsen} are not
appropriate for describing the low mass strangelets of interest here as
shell effects get crucial for their stability.  Similar shell model
calculations have been performed recently \cite{gils93}. The authors
consider strong {\em and} weak  decay by nucleon and hyperon emission
together  and conclude that stable low mass strangelets ($A_B<100$)
exist which have a low (and positive)  charge to mass ratio. A detailed
inspection of Fig.~2 in \cite{gils93} shows that strangelets with mass
number $A_B=10$ and $Z=-4$ or $Z=7$  exist which have a rather high mass
to charge ratio contrary to their conclusion.  We point out that the
procedure of their work is to start in the {\em absolute minimum}  for a given
baryon number and then to look for its stability against
possible weak and strong
decays.  Although
this investigation is very important, we believe that their finding is
not fully consistent with the initial condition of possible strangelet
production in relativistic heavy ion collisions. Initially, all kinds of
strangelets might be produced with different quark contents. They are
decaying first by strong decays (strong hadron emission) and afterwards,
if surviving, by weak hadronic decays. In Ref.\ \cite{gils93} the authors
start only with those strangelets at its minimum value in $E/A$ for a
given baryon number $A$ and study the stability of those candidates only.
However, it is only the slowest decay, the weak leptonic decay, which
ultimately drives a strangelet to its minimum value. For the purpose of
detecting strangelets the direct strong decays and weak hadronic decays
are now relevant, thus giving raise to a much wider class of possible
candidates to be observed (see also the discussion below).
In the following we recalculate strangelets for low masses and check now
separately for strong and weak hadronic decay starting with all possible
combinations and allowing for all kind of decays. 

The MIT bag model used includes only the quark kinetic energy, the Pauli
principle and confinement. 
It should be not taken too seriously as it effectively
models the features of QCD at the confinement energy scale.
It can not fix
the overall energy scale due to the unambiguity of the bag constant $B$,
which decides whether or
not strange quark matter is stable.
It does, however, illustrate potentially interesting effects as shell
closure effects.
Infinite strange quark matter, treated as a gas of noninteracting quarks,
is absolutely stable
for bag constants of $B^{1/4}\approx 145$ MeV.
For bag constants between
$B^{1/4}\approx 150-200$ MeV
strange quark matter is metastable, i.e.\ it can decay weakly 
depending on its strangeness content.
For larger bag constants strange quark matter is unstable and decays
completely via strong reactions into hadrons. Hence, we will discuss
our results for various choices of the bag constant, preferably for the
case of metastability.

The Dirac equation with a
Bogolyubov--type boundary condition reads
\begin{equation}
j_{l_\kappa}(pR)=-{\rm sgn}(\kappa)\frac{p}{E+m_i}j_{l_{-\kappa}}(pR)
\end{equation}
with
\begin{equation}
p=\frac{\omega_{\kappa,\alpha}^i}{R}
\end{equation}
and the energy
\begin{equation}
E_{\kappa,\alpha}^i=\left(\left(\frac{\omega_{\kappa,\alpha}^i}{R}\right)^2+
m_i^2\right)^{1/2}\; .
\end{equation}
Here $m_i$ denotes the mass of the quark of flavour $i$.
We choose $m_u=m_d=0$ MeV
and $m_s=150$ MeV. $R$ is the radius of the bag, $\kappa$ the
angular momentum quantum number and
$\alpha$ labels the eigenvalues in the quantum
state $\kappa$. The total energy is calculated by summing the lowest
occupied single particle energies and adding the phenomenological bag
energy $BV$ with the volume of the bag $V$, which has to be chosen such
that the inside pressure of the quarks is in equilibrium with the
outside vacuum pressure. The term
\begin{equation}
E_{\rm Cb} = \frac{1}{15}\alpha\frac{\left(2A_u-A_d-A_s\right)^2}{R}
\end{equation}
with the quark number $A_i$ of flavour $i$ accounts for Coulomb
corrections. Pressure equilibrium is achieved by minimizing the total
energy with respect to the bag radius. This
exact result has recently been approximated by
curvature contributions to the mass of
strangelets \cite{madsen}.

We calculate the binding energy of strangelets for an {\em arbitrary} number
of $u$--, $d$--, and $s$--quarks with $A_B\le 16$ for a given bag parameter.
Afterwards we look for possible strong decays, i.e.\ single baryon
\{n,p,$\Lambda,\Sigma^{+,-},\Xi^{0,-},\Omega^-\}$ emission and
mesonic $\{\pi^{+,-},K^{+,-},K^0,\bar K^0\}$ decays (note that kaon
decays do not occur in our calculation) extending the ideas already
presented in \cite{chin} to finite size configuration. Note that for
smaller strangelets shell effects are therefore of crucial importance.
For example a strong neutron decay of a strangelet
\begin{equation}
\label{strongnd}
Q(A_B,S,Z) \to Q(A_B-1,S,Z) + n 
\end{equation}
is allowed if the energy balance of the
corresponding reaction is
\begin{equation}
E(A_B,S,Z) > E'(A_B-1,S,Z) + m_n 
\end{equation}
where $E$ stands for the total energy of a strangelet.
Multiple hadron emission
is implemented by considering the combination of hadrons for a given charge
$f_z=Z/A_B$ and strangeness fraction $f_s=|S|/A_B$ with the lowest total
mass. The possible area for strangelets ($f_s \ge 0$, $f_z \ge -1$, and
$f_s+f_z\le 2$) is divided into 9 areas of (free) hadronic matter
where 7 areas are covered by
the possible metastable combinations of three different baryon species
as given in Table~4 of Ref.\ \cite{scha92}. 
The remaining two are $\{\pi^+p\Sigma^+\}$ and $\{\pi^-n\Sigma^-\}$ matter.
Multiple hadron emission is allowed if
\begin{equation}
E(f_s,f_z) > H(f_s,f_z)
\end{equation}
where $H(f_s,f_z)$ stands for the lowest mass of hadrons for a given
strangeness and charge fraction. Note that the baryon number does not
enter here as it is a conserved quantity.
In addition, we checked also
for fission of a strangelet into another strangelet and an arbitrary
number of hadrons while conserving charge, strangeness and baryon number
\begin{equation}
E(A_B,S,Z) > E'(A_B',S',Z') + (A_B-A_B')\cdot H(f_s',f_z')
\end{equation}
by three combined loops where $f_s'=(S-S')/(A_B-A_B')$
and $f_z'=(Z-Z')/(A_B-A_B')$. This allows for example for a combined
strong decay of a strangelet emitting a neutron and a pion. It might
well be that single hadron decay is not possible while multiple hadron
decay is due to shell effects.

Note that this procedure is different from the one used in \cite{gils93}
where mesonic decays and multiple hadron emissions have not been
implemented.
Nevertheless, the authors allowed also for weak neutron decay which we will
discuss later in a wider class (section \ref{sec5}) separately. 

A strangelet is called {\em metastable} in the following
if its energy lies under the corresponding (free) hadronic matter
of the same baryon number, charge, and strangeness, and if it can not
emit a single hadron or multiple hadrons by strong processes as
described above.
A metastable strangelet can then only 
decay via weak decays like the nonleptonic
(hadronic) decays.
Strangelets which are stable against strong decay but unstable against
nonleptonic weak decay
will be denoted in the following as {\em short--lived} (see section
\ref{sec4}).
Estimates of this weak decay rate range from
$\tau_{nl}\approx 10^{-6}-10^{-5}$ s \cite{koch} and
$\tau_{nl}\approx 3\times 10^{-7}$ s \cite{henning} to anything
between
$\tau_{nl} = 10^{-5}-10^{-10}$ s \cite{madsenweak}.
These estimates were calculated for the process 
$u+s\leftrightarrow d+s$ in infinite matter and depend sensitively on
the difference of the $d$-- and $s$--quark chemical potentials.
For the lightest strangelet, the H--dibaryon with quark content
($uuddss$), one gets a lifetime of about $\tau_{nl} = 10^{-8}-10^{-6}$ s
depending on its mass \cite{don}. 

Weak nonleptonic or hadronic decay is implemented by the hadronic decay
processes mentioned above allowing for strangeness violation 
of $\Delta S=\pm 1$. 
A strangelet stable against all these weak decays can
then only decay via leptonic decay of the form 
$s\leftrightarrow u{\rm e}^-\bar \nu_e$ and
$d\leftrightarrow u{\rm e}^-\bar \nu_e$ or via radiative decays 
through $us\leftrightarrow ud\gamma$.
These decay modes are suppressed by the three--body space of the
leptonic decay or by the electromagnetic coupling constant 
of the radiative decay compared to the nonleptonic decays.
Both decay modes are supposed to yield similar lifetimes 
\cite{berger} which are then 
higher than the nonleptonic ones.
A strangelet stable with respect to weak hadronic decay but unstable with
respect to weak leptonic or radiative decay
is called {\em long--lived} (see section \ref{sec5}). It lives 
on the time scale of the weak leptonic decay $\tau=\tau_{l}$,
which has been estimated to be
$\tau_l = 10^{-4}-10^{-5}$ s in infinite matter
\cite{chin,hhleptonic,koch}. 
If a strangelet demonstrates to be stable against these decays also, then
it can only decay by higher order weak decay 
($\Delta S=\pm 2$), which results in lifetimes of days
(see e.g.\ the estimate of the $\Delta S=\pm 2$ decay of the H--dibaryon
\cite{don}). In this case the strangelet would be super long--lived.

\section{MEMOs}
\label{sec3}

The (multi-) strange baryonic objects are treated within the framework of an
extended relativistic
mean--field theory. Although the application of the mean--field
approximation seems doubtful for small systems, it has been shown that the
model furnishes quite remarkable results for nuclei with baryon numbers
as small as $A_B=4$ \cite{scha92}. In addition to the well--known $\sigma$--
and $\omega$--meson, strange scalar and vector mesons are introduced, the
$\sigma^*$-- and the $\phi$--meson ($m_{\sigma^*}=975$ MeV,
$m_\phi=1020$ MeV). The latter couple to strangeness only, thus
incorporating the seemingly strong
attractive hyperon--hyperon interaction \cite{scha93}.

The vector coupling constants are
chosen according to SU(6)--symmetry, the scalar coupling constants are
fixed to hypernuclear data.
Within this model, 
MEMOs consisting of
combinations of $\{p,n,\Lambda,\Xi^0,\Xi^-\}$ baryons
demonstrate to be metastable due to Pauli--blocking effects. 
They possess binding
energies per baryon of $E_B/A_B\approx -22$ MeV, strangeness per
baryon of up to $f_s\approx 2$, unusual charge per baryon of 
$f_z\approx -0.5$ to zero while carrying positive baryon number and
baryon densities up to $2.5 - 3$
times that of ordinary nuclei.
Metastable clusters of purely hyperonic matter
$\{\Lambda,\Xi^0,\Xi^-\}$ have been also predicted \cite{scha93}.

In the following we extend the calculation of \cite{scha93} to small
mass numbers relevant for heavy ion physics.
We calculate even combinations of
\{n,p,$\Lambda$,$\Xi^0$,$\Xi^-$\} baryons up to a filled s-- and p--shell, 
i.e.\ 8 baryons of each
using model 2 of Ref.\ \cite{scha93}.
Out of these 3125 combinations we have found 298 configurations,
which are bound and metastable.
The two smallest systems for $A_B=4$ are
$^4$He and the corresponding $\Xi$--system, i.e.\ two $\Xi^-$ and two $\Xi^0$.
The next heavier ones ($A_B=6$) are the combinations
$_{\Lambda\Lambda}^{~~6}$He, $\{2{\rm n},2\Lambda,2\Xi^-\}$,
$\{2{\rm p},2\Lambda,2\Xi^0\}$, and $\{2\Lambda,2\Xi^0,2\Xi^-\}$.
Pure $\Lambda$ or neutron matter is not bound in this model.
Nevertheless, we have found some very loosely bound $\{\Xi^0,\Xi^-\}$--systems.
The properties of these small MEMOs are summarized in
Table~\ref{tab:memos}.
Note that the double $\Lambda$ hypernucleus $_{\Lambda\Lambda}^{~~6}$He
has already been seen \cite{Prowse}. 
Other light candidates are discussed extensively in \cite{scha93}, like 
$^{~~~~6}_{\Xi^0\Xi^0}$He and
$^{~~~~7}_{\Lambda\Lambda\Xi^0}$He.

The binding energy per baryon for these small systems
is not more than $-16$ MeV for baryon numbers less than $A_B\approx 16$.
For higher baryon numbers we have found an approximately
linear decrease of the binding energy. This fact is already known from
ordinary nuclei. The less bound combinations found are mainly combinations
of $\Xi^-$ and $\Xi^0$, i.e.\ pure $\Xi$--matter with only 
$E_B/A_B\approx -2$ MeV.

MEMOs decay by weak mesonic or nonmesonic decay in analogy to
hypernuclear weak decay. The nonmesonic decays of the type
$\Lambda N\to NN$, $\Xi N,\Lambda\Lambda\to\Lambda N,\Sigma N$, $\dots$
will play an important r\^ole \cite{scha93}. Due to the high mass
differences involved in these decays (about $180-200$ MeV) the process
is not hindered by Pauli--blocking effects. 
The first process has been seen in 
the weak decay of hypernuclei, 
which decay on the time scale of the lifetime of the $\Lambda$, i.e.\
$\tau_w\approx\tau_\Lambda\approx 10^{-10}$ s. 
Also the mesonic decay $\Lambda\to N\pi$ yields similar lifetimes for
very light hypernuclei (see e.g.\ \cite{hyper}).
We expect that MEMOs will live then on the same time scale, 
$\tau_{\rm MEMO}\approx \tau_w$,
irrespective of their strangeness content, if they are not deeper
bound to create a minimum in the {\em total} energy at finite
strangeness as it is the case for strangelets.

\section{Short--lived Strange Matter}
\label{sec4}

In the following we will discuss the properties of light strangelets
and compare them to those of MEMOs.

\begin{figure}
\epsfxsize=\textwidth
\epsfbox{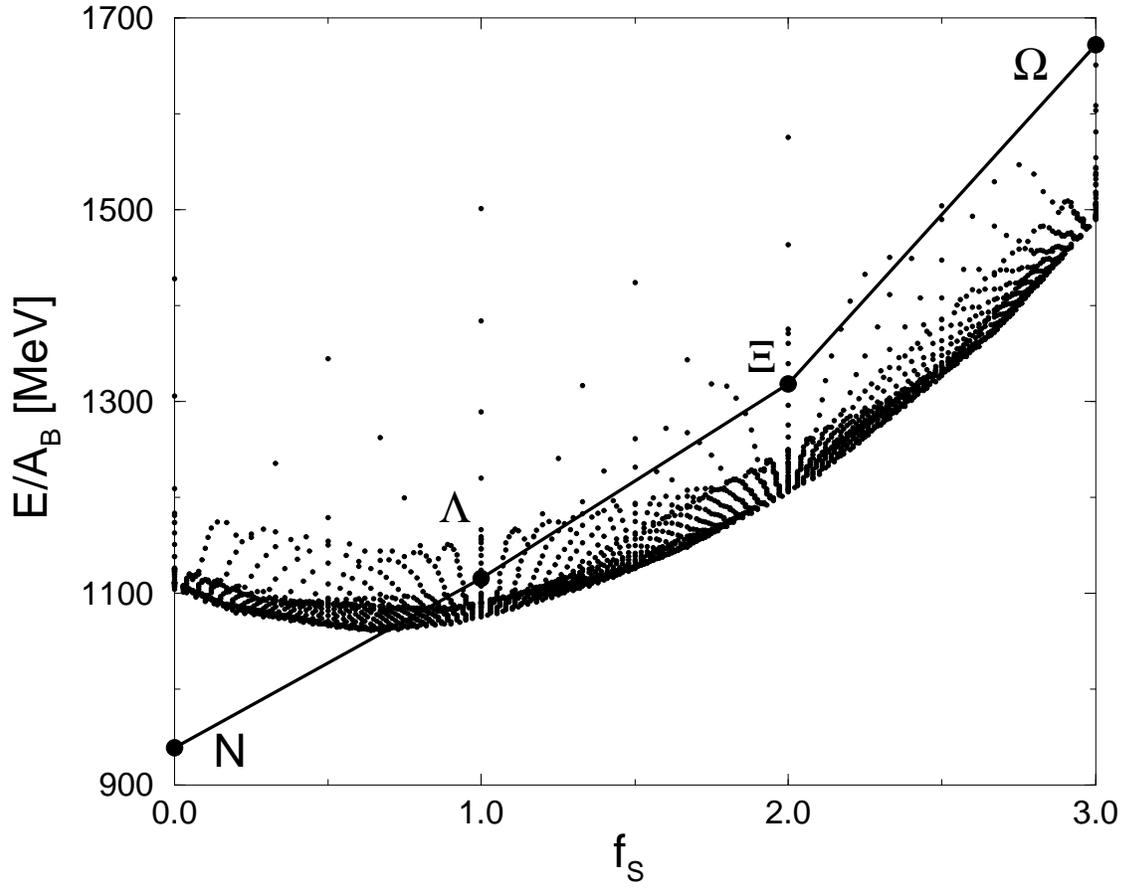}
\caption{The energy per baryon $E/A_B$ of all possible strangelets
with $A_B\le40$ and
$N_u=N_d$ for a bag constant of $B^{1/4}=170$ MeV versus
the strangeness fraction $f_s$. The solid line
connects the masses of nucleon, $\Lambda$, $\Xi$ and $\Omega$ and stands
for free baryon matter.}
\label{fig1}
\end{figure}

Fig.\ \ref{fig1} shows the energy
per baryon of all possible quark bags with a baryon number up to $A_B=40$
as a function of the strangeness per baryon $f_s = |S|/A_B$ for a bag
constant of $B^{1/4}=170$ MeV.
We only show bags with equal numbers of up and down quarks. 
The energetically most favourable combinations always
have the same number of up and down quarks since they occupy the same
single--particle levels. The Coulomb correction is in the order of some
MeV and thus not important in our case.

The solid line connects the masses
of the nucleon, $\Lambda$, $\Xi$ and $\Omega$. 
As a first cut for potential candidates, quark bags lying above
this line can (and probably will) {\em completely} decay via strong processes,
those lying beneath the line
will lead in principle to metastable strangelets. 
The important point to note here is
that any strangelet initially formed
with a mass under this line might decay to
another strangelet changing baryon number, strangeness and charge, but 
can not decay to a pure hadronic state anymore, simply because of energy
conservation! 
It is remarkable that there exists
a quite sharp lower limit of the binding energies:
already for bags with $A_B\approx 40$
infinite quark matter is quite a
good approximation. 

It is clear that this simple version of the MIT--Bag
model is not able to reproduce the hadron spectrum. The masses of the
nucleon, $\Lambda$ and $\Xi$ are overestimated, the mass of the $\Omega$
is underestimated. To provide a better description of the hadron masses,
one--gluon exchange corrections and a zero--point energy term of the form
$Z_0/R$ must be taken into account 
\cite{degr75}. Strangelets with s--states only 
up to a mass number of $A_B=6$ have been studied
extensively in \cite{aerts} taking into account colourmagnetic and
colourelectric corrections. No strangelet has been found to be metastable
with the exception of the H--dibaryon with $A_B=2$, $S=-2$ and $Z=0$. 
This finding has been confirmed in \cite{farhi}. Up to now, these
corrections can only be applied to quarks sitting in s--states. We are
not aware of any attempt including these terms for higher shell states
also. 

For a bag constant of $B^{1/4}=145$ MeV (not shown)
almost all bags are lighter than the corresponding hyperonic matter.
After subsequent strong and weak processes the final resulting strangelets
thus should have long enough lifetimes to be detectable. 
Nevertheless, for the situation depicted in Fig.\ \ref{fig1}
($B^{1/4}=170$ MeV)
strangelets with strangeness
fraction less than $f_s\approx 0.6$ are less bound than corresponding
hyperonic matter (solid line) and hence, they can completely decay strongly into
hyperons and nucleons and are
not detectable in heavy ion searches.
For intermediate strangeness fractions only bags with rather
large baryon number then can be stable with respect to strong decay. 
This means that there exists a minimum critical strangeness fraction 
for which strangelets exist which will depend of course on the bag
parameter and baryon number of the strangelet.

The dependence on the baryon number and the bag parameter is
illustrated in the following. 
We have calculated the mass of all possible bags with $A_B\le 16$,
which are most interesting for heavy ion experiments,
but with arbitrary numbers of up, down and
strange quarks and various bag parameters.
As already stated, we define metastable
strangelets as those who are stable against strong decay, while unstable
strangelets are not.

\begin{figure}
\epsfbox{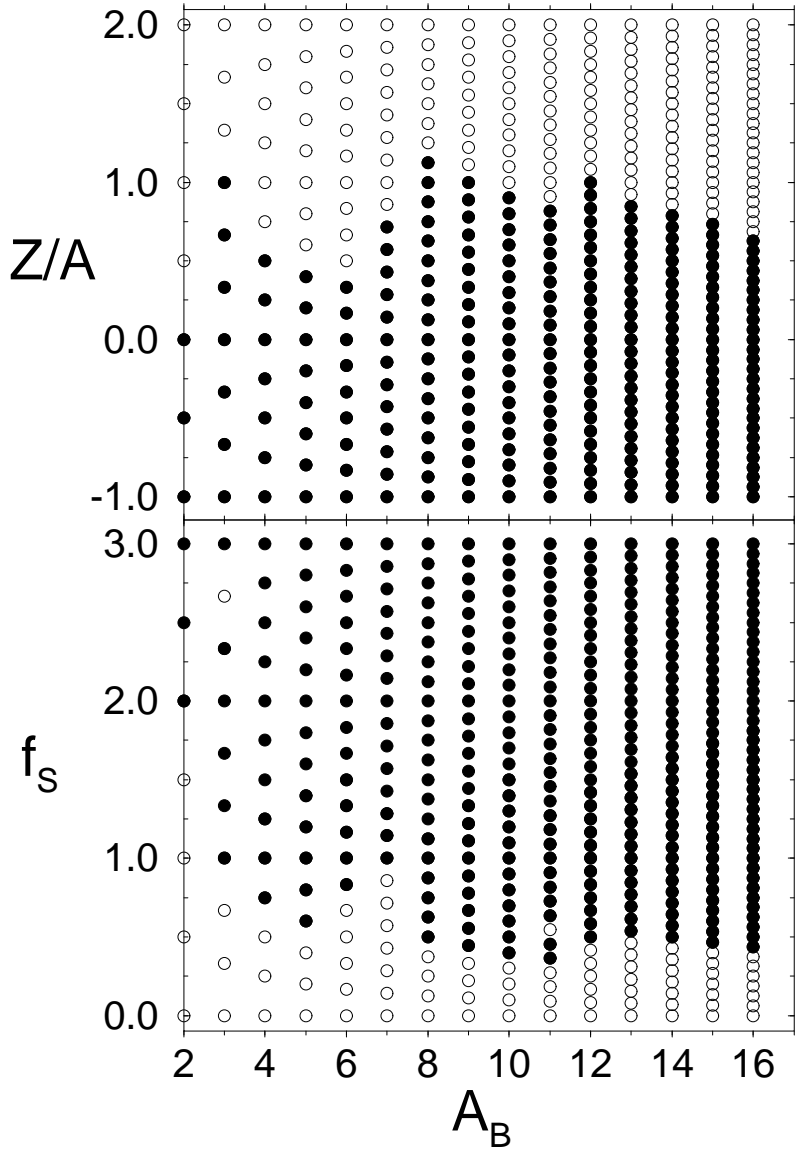}
\caption{The strangeness per baryon $f_s$ (lower part) and the charge
fraction $Z/A$ (upper part) as a function of the baryon
number $A_B$ for short--lived strangelets (dots), 
unstable strangelets (open circles) for a bag constant of
$B^{1/4}=150$ MeV. The hadronic counterpart, MEMOs, are shown by crosses.}
\label{fig150}
\end{figure}

\begin{figure}
\epsfbox{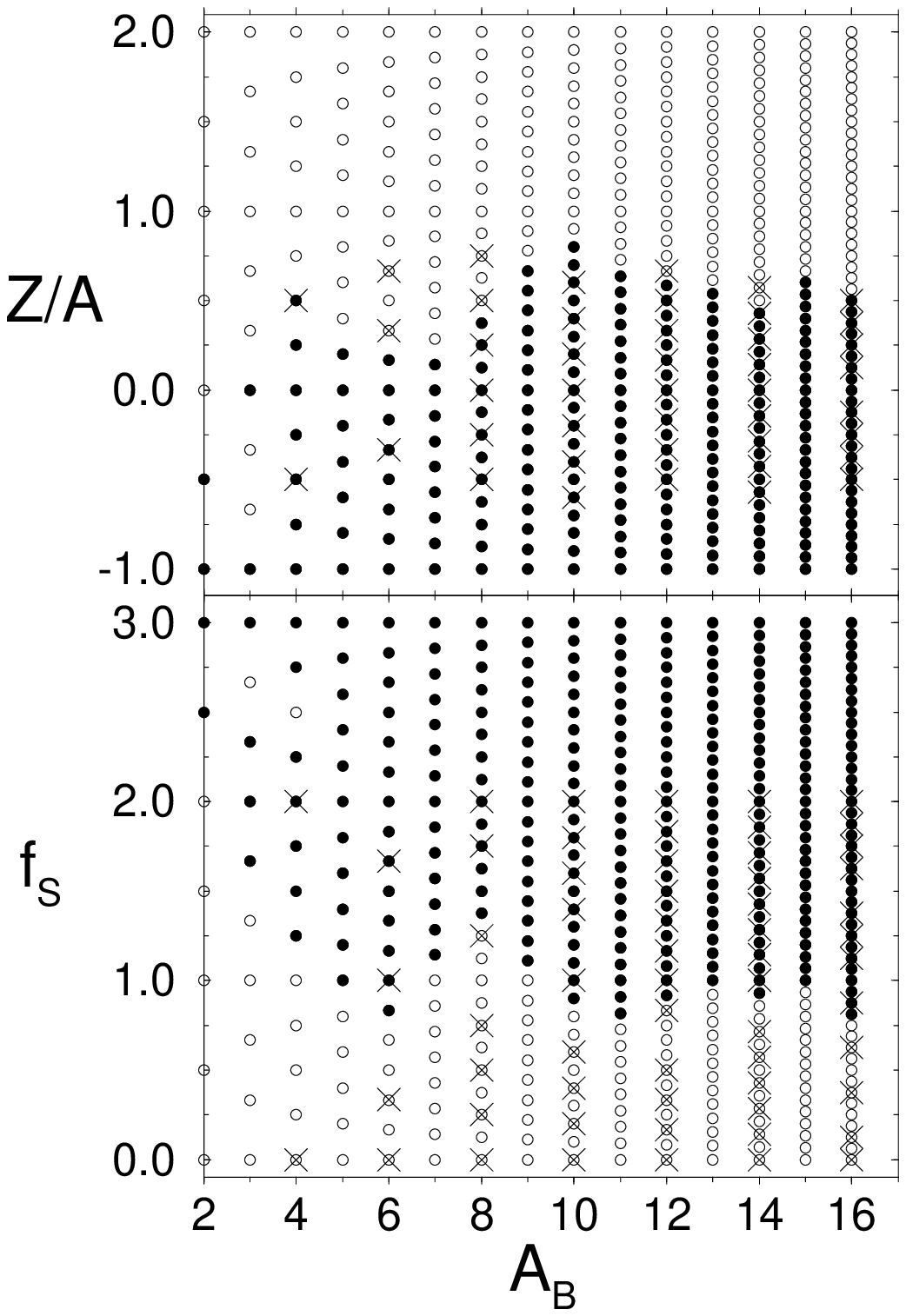}
\caption{The same as Fig.\ \protect\ref{fig150} for
a bag constant of $B^{1/4}=160$ MeV.}
\label{fig160}
\end{figure}

\begin{figure}
\epsfbox{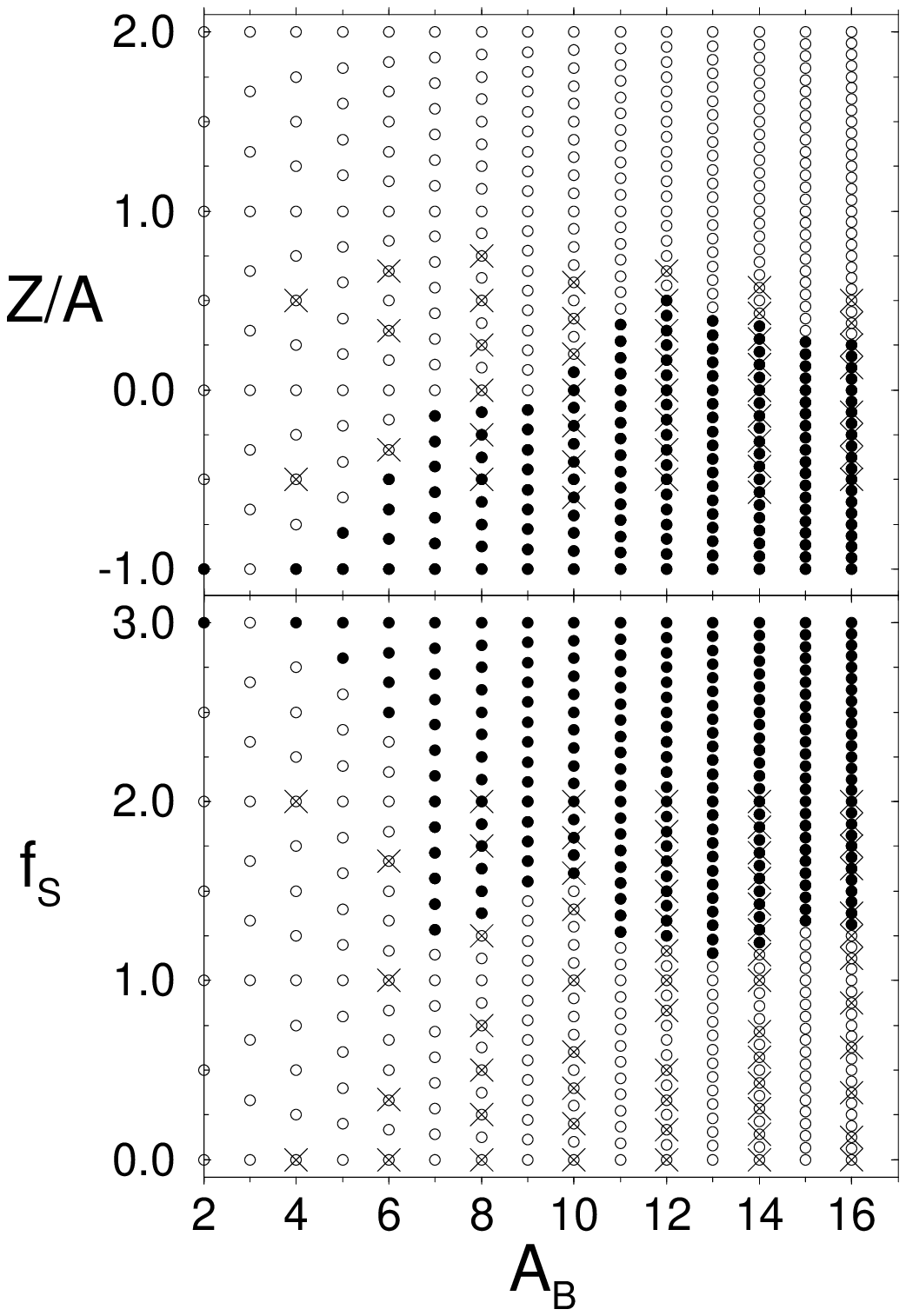}
\caption{The same as Fig.\ \protect\ref{fig150} for
a bag constant of $B^{1/4}=170$ MeV.}
\label{fig170}
\end{figure}

Figs.\ \ref{fig150}--\ref{fig170} show the strangeness and the charge fraction
versus the baryon number for bag parameters of $B^{1/4}=150,160,170$
MeV, respectively.
Dots stand for metastable, 
open circles for unstable bags and crosses for small MEMOs.
For higher bag parameters, one sees less candidates of metastable
strangelets and they are shifted to higher strangeness, higher masses 
and, more important, to negative charges!

For $B^{1/4}=150$ MeV shown in Fig.\ \ref{fig150}, metastable
strangelets exist for a wide range of charge ($|Z|/A\leq1$)
and strangeness fraction. Only for quite low strangeness fraction
$f_s<0.4$, comparable to the ones of light double $\Lambda$ hypernuclei, 
strangelets are unstable with
respect to strong decay.
This situation changes when looking at 
the case of $B^{1/4}=160$ MeV depicted in Fig.\ \ref{fig160}.
Nearly no strangelet appears to be metastable 
with $f_s<1$ or $Z/A> 0.6$ in the mass range considered here
as they are subject to fissioning into nucleons and hyperons. 
Especially for light systems, the maximum
charge is $Z=+2$. This trend is getting even more pronounced for 
$B^{1/4}=170$ MeV as can be seen in Fig.\ \ref{fig170}.
Only a few of the very light strangelets for $A\leq 6$ remain to be
metastable. They are highly negatively charged $Z/A_B<-0.5$ and have a 
very high strangeness fraction of $f_s>2.5$.
For higher baryon numbers ($6<A_B<16$)
no metastable strangelet exists for $f_s<1$ and $Z/A>0.5$.
Most of the metastable strangelets are found to be
highly negatively charged which is
contrary to the conclusion drawn in \cite{gils93}.
Here the authors start their consideration in the minimum for a given
baryon number due to flavor equilibrium. 
Nevertheless, a hot strangelet formed in heavy ion collision does not
start decaying with flavor
changing decays but with strong decays.
Our results demonstrate that the cascade of strong decays
of a strangelet do not stop
in an absolute minimum but 
in a region allowing for highly charged strangelets.

The hadronic counterpart, MEMOs, are also shown as crosses
in figs.\ \ref{fig150}--\ref{fig170} for comparison.
There are candidates which are highly
positive and negative charged ($-0.6<Z/A<+0.7$).
MEMOs show up where strangelets are unstable and vice versa.
There is also a region in the $f_s-Z/A$ plane where both 
MEMOs and strangelets appear.
Here, the energetically least favourable object
will decay into the other.
A strangelet created
in a quark--gluon plasma can then possibly decay into a MEMO via strong
interactions. 
On the other side, MEMOs can coalesce from the hot and hyperon--rich
zone of a relativistic heavy ion collision first and form
a strangelet which is then detected.
The density distributions within both objects resemble each other
closely.
Transition matrix elements are to first
approximation proportional to the overlap of both wavefunctions.
Therefore, the energetically least favourable state may be a doorway state,
decaying rapidly into the favourable state.
In \cite{kerb} the decay for a double hypernucleus to a
H--dibaryon was estimated to happen at $\tau=10^{-18}-10^{-20}$ s if
the masses of the H--dibaryon is close to $2m_\Lambda$. We expect
similar time scales for the transition of one form of strange matter to
the other if the masses are similar.
Of course, this is speculation and can not be clarified by our present
knowledge of strange matter.

\section{Long--lived Strange Matter}
\label{sec5}

The properties of strange matter as discussed in the previous section
apply for systems living on the time scale of weak interactions, i.e.\
$\tau_{\rm MEMO}\approx 10^{-10}$ s and $\tau_{nl}=10^{-5}-10^{-10}$ s.
Present experiments are looking for long--lived strangelets 
$\tau_{\rm exp} > 10^{-6}-10^{-8}$ s. 
They are not able to see MEMOs and
possibly most of the strangelet candidates, 
if the nonleptonic decay is too fast. Nevertheless, it is known that
the leptonic decay gives longer lifetimes
due to the reduced
phase space of the three body decay \cite{chin,hhleptonic,koch}. 
As $\tau_l = 10^{-4}-10^{-5}$ s in infinite matter, the present
experiments are sensitive to strangelets which are stable against
weak nonleptonic decay.
In the following we study
all possible weak hadronic decay for the metastable strangelets, 
as weak pion, kaon, proton, neutron, $\Lambda$, $\Sigma^{+,-}$, $\Xi^{0,-}$, 
$\Omega^-$ decays.

For example a weak neutron decay of a strangelet (which turns out to
be one of the most dominant decay modes)
\begin{equation}
\label{weaknd}
Q(A_B,S,Z) \to Q(A_B-1,S-1,Z) + n 
\end{equation}
happens if the energy balance of the
corresponding reaction reads
\begin{equation}
E(A_B,S,Z) > E'(A_B-1,S-1,Z) + m_n 
\end{equation}
where $E$ stands for the total energy of a strangelet.
Note that weak neutron decay drives a strangelet to lower strangeness
fraction only if the initial strangeness content of the strangelet is $f_s<1$,
but to {\em higher} strangeness fraction for
$f_s>1$ as the obtained shift in the strangeness content is easily found to
read $\Delta f_s = (f_s-1)/(A-1)$ and thus gets positive in this case!

Multiple hadron emission and fission to another strangelet is also
checked in analogy to the strong decays described in section \ref{sec2}
but now with $\Delta S=\pm 1$.
In addition we add the case with
$B^{1/4}=145$ MeV, $m_s=280$ MeV which are the original MIT values
\cite{degr75}. 

\begin{figure}
\epsfxsize=\textwidth
\epsfbox{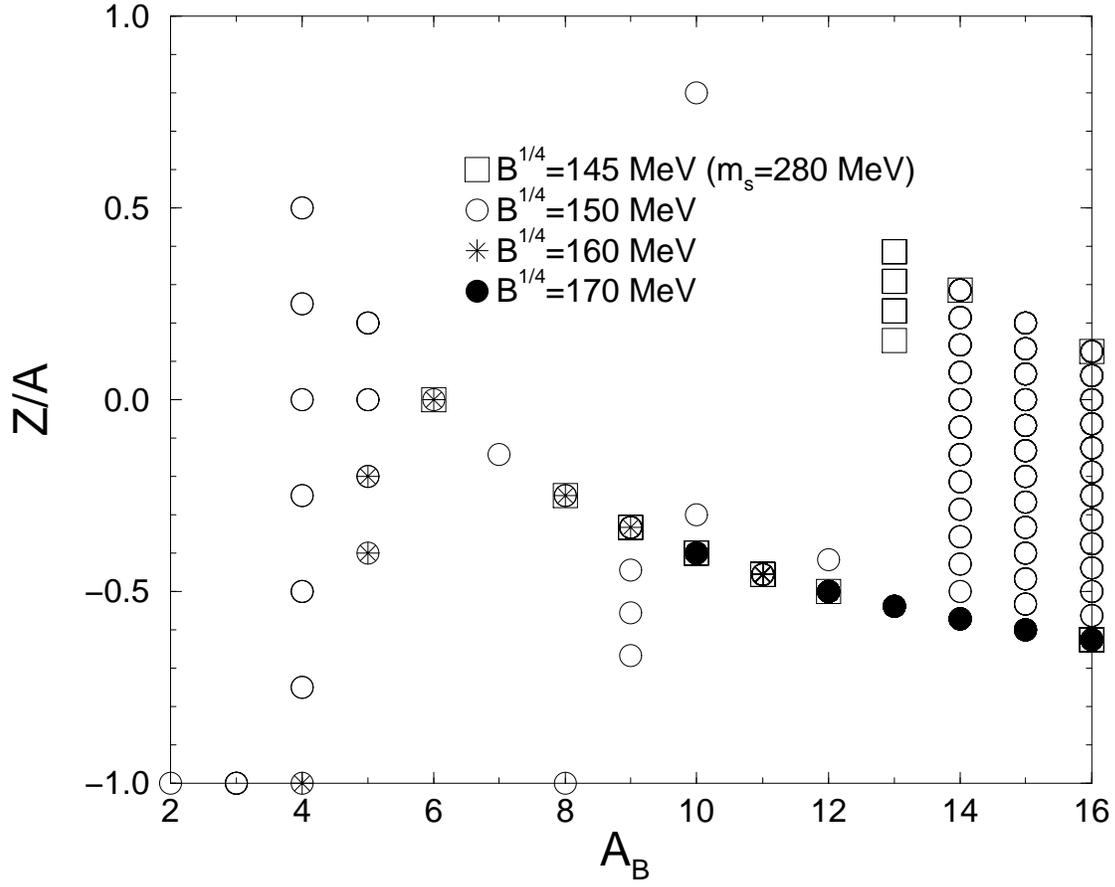}
\caption{The charge fraction $Z/A$ 
for long--lived strangelets, which are stable against 
nonleptonic weak decay, for different choices of the bag parameter.
The case for the original MIT parameters ($B^{1/4}=145$ MeV, $m_s=280$
MeV) is also shown.}
\label{figlong}
\end{figure}

We find that the strangelets mainly decay via weak pion or baryon
emission. In a very few cases weak multiple hadron emission is possible.
We have also checked the case $B^{1/4}=180$ MeV but none of the
strangelets are stable with respect to weak hadronic decay.
Nevertheless, for the other cases we found some candidates.
The remaining long--lived strangelets are shown in Fig.\
\ref{figlong}. Surprisingly, we find some long--lived strangelets with
quite low mass numbers for all cases considered here. 
They are lying on a chain which starts from the triple
magic strangelet ($6u6d6s$) where all quarks fill up the 1s--state --
due to its symmetry character it is also called the quark alpha
\cite{qalpha}. The 'valley of stability' starting at the quark alpha
continues then towards negative charges by adding one unit of negative
charge when going to a higher mass number. The reason for the
stability line is a pronounced shell effect. 
These strangelets mainly have a closed s--shell for
the $u$--quarks and a closed s--, p3/2-- or p1/2--shell 
for the $s$--quarks. 
Then the $d$--quarks added result in the chain seen in Fig.\
\ref{figlong}. This rule is less stringent for the case
$B^{1/4}=150$ MeV as many strangelets demonstrate to be long--lived.
For the case $B^{1/4}=145$ MeV also strangelets with a
closed p3/2--shell for the $u$--quarks appear resulting in the
positively charged candidates at $A_B=13,14$. Due to the higher strange
quark mass of $m_s=280$ MeV strangelets with high strangeness fraction
are less stable and only the 1s--shell for the $s$--quarks are filled
here. Also for $B^{1/4}=150$ MeV there are several cases
with a filled p3/2--shell for the $u$--quarks, 
like $A=10$, $Z=+8$ ($18u6d6s$).
The candidates shown for $A_B\leq 6$ have to be taken with some care as
colourmagnetic and colourelectric terms are not included in the present
investigation. If included, they appear to be not stable at all
\cite{aerts,farhi} and can decay to ordinary hadrons via strong
interactions. The lightest long--lived candidates for $A_B\geq 6$ are 
summarized in Table~\ref{tab:longlived}. 
We add also the candidates which can only decay by weak multiple
hadron emission as the decay is suppressed by phase space. Actually,
we find only one additional strangelet of this type.

Most promising candidates are
for $A_B=10$, $Z=-4$, for $A_B=12$, $Z=-6$ and $A_B=16$, $Z=-10$, which
appear in all the cases studied here. Note that the first candidate is
triple magic ($6u6d18s$ or $6u18d6s$). 

The question arises, why mainly {\em negatively} charged strangelets appear to
be stable against weak hadronic decay (and also already, to a somewhat lesser
extent, against strong hadronic decay).
The reason is twofold: (a)
Strangelets with
a rather low strangeness fraction (and correspondingly positive charges)
are decaying strongly by fissioning into nucleons and hyperons
as discussed in the last section. As $Z/A>0$ follows model
independent from $f_s<1$ for isospin saturated matter this will always
happen if 
there is a minimum energy for strangelets at high strangeness fraction.
Still, in addition, any strong decay by emitting a neutron (cf.\
(\ref{strongnd}))
(or a proton) will enhance the relative strangeness content of the remaining
strangelet by an amount $\Delta f_s = f_s /(A-1) >0$ shifting it to a higher
strangeness fraction $f_s$ which thus can exceed 1.
(b) It is easier for positively
charged strangelets to decay via e.g.\ weak proton decay. The charge reverse
reaction would be a weak neutron decay accompanied by $\pi^-$ emission
which is less favourable. In general, (the dominant) weak nucleon decay
(cf. (\ref{weaknd}))
will drive strangelets with $f_s>1$ to a higher strangeness
fraction and hence to higher negative $Z/A$-ratios.

These two reasonings should be generally valid if the masses of
finite droplets of SQM follow a distribution closely to that shown
in Fig. \ref{fig1}. 
Although calculated within the MIT bag model, we believe that
a similar distribution would in principle also show up when applying
other bag models (and fixing the same overall energy
scale within the appropriate parameters). In this sense our major result,
i.e. the tendency of short-lived and long-lived metastable strangelets
to exist preferably as a slightly or highly negative exotic state of matter,
should be seen to be valid on more general grounds.

We want to stress again that the cascade of weak decays of a strangelet
is trapped in a deep local minimum where it can only decay via weak
semileptonic and radiative decay.
We have checked the stability
with respect to these two decays.
In only two cases a
strangelet demonstrates to be stable against these decays:
this is the quark alpha
for $B^{1/4}=150$ MeV and $B^{1/4}=160$ MeV, and the strangelet
($6u6d3s$) with $A_B=5$, $Z=+1$ for $B^{1/4}=150$ MeV.
They would live on the time scale of days as only weak decays with
$\Delta S= \pm 2$ are allowed. Due to its nonzero charge only the later
would be visible in
heavy ion experiment.
Nevertheless, the colourmagnetic term is
repulsive for these multiquark states. For the quark alpha, the 
colourmagnetic term is about +150 MeV/A without symmetry breaking
effects, i.e.\ half the mass splitting of the nucleon and Delta. 
Therefore, the colourmagnetic interaction would shift the mass above the
mass of 6$\Lambda$'s. Then, the quark alpha is simply a resonance.

It is interesting to study now the detectability 
of these long--lived strangelets in heavy ion collisions 
especially for $A_B>6$.
All recent experiments searching for strangelets 
are sensitive to composites with lifetimes of
$\tau>10-100$ ns except for experiment NA52 with $\tau>10^{-6}$ s.

Experiment E814 \cite{barr90} was looking for positively charged
strangelets with $0.1<Z/A<0.3$, $A_B>10$.
Their limit on the production of
a strangelet in a single event was $1.2\times 10^{-4}$ for multiply
charged strangelets.
Experiment E886 was set up to
$0<A/Z<14$ yielding a 
much higher sensitivity of down to $10^{-7}$/event \cite{E886}.
An open geometry is used by experiment E864 which reported
new limits for $Z=+1,+2$ and a wide range of mass $A_B>10$ 
of about $10^{-5}-10^{-6}$/event 
just recently \cite{E864}. 
According to Fig.\ \ref{figlong} these experiments 
could see the positively charged
candidates for $B^{1/4}=145$ MeV ($m_s=280$ MeV)
at $A_B=13,14,16$ and the ones
for $B^{1/4}=150$ MeV at $A_B=14-16$. 

The high sensitivity experiment E858 
\cite{aoki92} was looking in the range $-1<A/Z<-7$ 
and no strangelet was found with a sensitivity of
$10^{-9}-10^{-10}$/event. Also the follow--up experiment 
E878 \cite{E878} using the gold beam did not see any evidence for
unusual composites with $|Z|\leq 3$ on the level of
$10^{-7}$/event. Nevertheless,
these experiments were measuring at zero degree and were not measuring
at midrapidity for $A_B>8$ \cite{E878}. 
Hence, they are only sensitive
to e.g.\ the candidate $A_B=7$, $Z=-1$ for the case $B^{1/4}=150$ MeV
and for the other charged candidates at $A_B<6$ for $B^{1/4}=150,160$
MeV. 

Unfortunately, none of these experiments has set limits so far for the
other candidates in the valley of stability, like 
$A_B=10$ and $Z=-4$ and for higher negative charges.
While finishing this work, new results from E886 were published which
give limits for negatively charged strangelets down to
$10^{-8}$/event, but unfortunately for $|Z|<4$ only \cite{E886}.
Most recently, experiment NA52 presented new limits for negatively
charged strangelets with a sensitivity down to $10^{-8}-10^{-10}$/event for
$M/|Z|=10-40$ \cite{bore94}. 
However, their limit for the above mentioned
strangelets with $M/|Z|\approx 2$ 
is much less, about $10^{-6}-10^{-7}$/event. 
Moreover, results were presented for rigidities of $p/Z=40,100,200$
GeV only, while NA52 can cover the range of $p/Z=5-200$ GeV.
Midrapidity is reached
for $M/Z\approx p/9Z$ \cite{bore94} corresponding to $M/Z=4.4$ GeV for
the lowest rigidity of $p/Z=40$ GeV measured so far. Lower rigidities
are therefore probing strangelets with lower $M/Z$ and NA52 would be
sensitive to the negatively charged strangelet candidates proposed here.

\section{Summary}

The present investigation seems to indicate that the search
for highly charged strange matter would be far more promising than
hitherto recognized.
Most search experiments are, however, designed
to detect only particles with small charge--to--mass ratio.
But long--lived positively charged strangelets
seems to exist only for $A_B>12$ and very low bag parameters.
The most interesting candidates for long--lived strangelets are lying in a
valley of stability which starts at the quark alpha ($6u6d6s$)
and continues by adding one unit of negative charge, i.e.\ 
$(A,Z)=(8,-2),(9,-3),(10,-4),(11,-5)\dots$.
The present experimental setups are hardly sensitive to these candidates.
Plans for extending experiment E864 to look
for highly charged strangelets for $A_B\geq 10$ are therefore most
interesting \cite{sandweiss}.

On the other hand, experiments looking for short--lived strange
matter will be able to see a much wider variety of combinations of
charge and mass. 
Recently, experiment E896 started looking into this rich domain of
short--lived composites for the H--dibaryon \cite{E896} but other
composites with low mass--to--charge ratios might be also accessible.
Short--lived strange matter, either in the form of
metastable strangelets or MEMOs, demonstrate to be also highly
negatively charged which opens the possibility for measuring their
formation with an extremely low background from antinuclei.
These metastable composites can be detected by a cascade of weak decays
and by their unusual charge and mass. 
Measuring single and double $\Lambda$ hypernuclei
will set limits on the production possibility of MEMOs.
This limit can possibly also be applied for
strangelets as MEMOs can decay to them via strong decay and 
serve as a doorway state. The other way around, 
if MEMOs are found they will set
stringent limits on the existence of strangelets for the same
charge, strangeness and mass and will give new impetus for 
our understanding of the strong interactions between baryons in general.

\acknowledgements

This paper is dedicated to our former collaborator, Carl B. Dover.
We thank H. Crawford, R. Klingenberg, 
J. Nagle, K. Pretzl, F. Rotondo, J. Sandweiss for useful
discussions and remarks. 
J.S.B. thanks the Niels Bohr Institute 
for the warm hospitality and the Alexander von Humboldt-Stiftung 
for its support with a Feodor-Lynen fellowship.
This work was supported in part by the Graduiertenkolleg
`Theoretische und Experimentelle Schwerionenphysik' of the Deutsche
Forschungsgemeinschaft (DFG), the Ge\-sellschaft f\"ur
Schwerionenforschung Darmstadt (GSI) and the Bundesministerium f\"ur
Bildung und Forschung (BMBF).

\begin{table}

\caption{
Some candidates for long--lived strangelets with $A\geq 6$ 
which are stable against weak nonleptonic decay.
Case I: $B^{1/4}=145$ MeV, $m_s=280$ MeV;
Case II: $B^{1/4}=150$ MeV, $m_s=150$ MeV;
Case III: $B^{1/4}=160$ MeV, $m_s=150$ MeV;
Case IV: $B^{1/4}=170$ MeV, $m_s=150$ MeV.
$^*$This candidate can decay via a collective nonleptonic weak decay.}

\begin{tabular}{*{14}{c}}
A & 6 & 7 & 8 & 8 & 9 & 9 & 10 & 10 & 11 & 12 & 12 & 13 & 13 \cr
\hline
Z & 0&--1&$-8,-3^*$&--2&$-4,-5,-6$&--3&--4&$-3,+8$&--5&--6&--5&--7&+2,3,4,5\cr
\hline
case & I-III & II & II & I-III & II & I-III & I-IV & II & I-III & I-IV & II
& II-IV & I \cr
\end{tabular}
\label{tab:longlived}
\end{table}

\begin{table}
\caption{Candidates of small MEMOs for filled s--shell states. Here
we only consider (pn$\Lambda\Xi^0\Xi^-$) baryons.
The double hypernucleus $^{~~6}_{\Lambda\Lambda}$He which was already
seen \protect\cite{Prowse} belongs also to this class. 
Candidates involving $\Sigma$ baryons
are 2(n$\Sigma^-\Xi^-$) and 2(p$\Sigma^+\Xi^0$).}
\begin{tabular}{*{10}{c}}
 & \hspace*{-15pt} 2($\Xi^0 \Xi^-$) 
 & 2($\Lambda \Xi^0 \Xi^-$) 
 & 2(n$\Lambda \Xi^-$)
 & 2(p$\Lambda \Xi^0$) 
 & 2(n$\Lambda \Xi^0 \Xi^-$) 
 & 2(p$\Lambda \Xi^0 \Xi^-$) 
 & 2(pn$\Lambda \Xi^-$)
 & 2(pn$\Lambda \Xi^0$) 
 & 2(pn$\Lambda \Xi^0 \Xi^-$) \cr
\hline
A & 4 & 6 & 6 & 6 & 8 & 8 & 8 & 8 & 10\cr
Z &--2&--2&--2&+2&--2&0&0&+2&0 \cr
S &8&10&6&6&10&10&6&6&10
\end{tabular}
\label{tab:memos}
\end{table}

\end{document}